# The Role of ESLEA in the development of eVLBI

R. E. Spencer[1], R. Hughes-Jones[1], M. Strong[1], S. Casey[1], A. Rushton[1], P. Burgess[1], S. Kershaw[1], C. Greenwood[2]

[1] School of Physics and Astronomy, The University of Manchester, Oxford Rd Manchester, M13 9PL, UK.
[2] National e-Science Centre (NESC), Edinburgh, EH8 9AA, UK

**Abstract**
The Internet has been used for data transfer in radio astronomy ever since its inception; however it is only recently that network bandwidth capability means that the Internet becomes competitive with traditional forms of data storage. Very Long Baseline Interferometry (VLBI) uses widely separated telescopes between which high bandwidth direct connections have not been feasible until recently. The academic networks now allow us to connect at high data rates (~1Gbps) in "eVLBI". The ESLEA project (Exploitation of Switched Lightpaths for E-science Applications) has played a major role in the development of eVLBI. We outline this development in this paper.

**Key words:**

VLBI, networks, lightpaths, TCP, UDP, Constant bit rate.

## 1. Introduction

VLBI (Very Long Baseline Interferometry) is an aperture synthesis technique that utilizes radio telescopes from around Europe and the rest of the world. The telescopes observe the same cosmic radio source simultaneously, and are equipped with accurate atomic clocks to maintain coherence. Signals from the telescopes are combined to produce images of celestial objects with high angular resolution. A typical experiment using European telescopes at an observing frequency of 5 GHz achieves a resolution of around 5 milli-arcseconds, equivalent to the size of a small house on the moon as seen from the Earth, and a factor of 10 better than that achieved by the Hubble Space Telescope. Traditionally the data were recorded on magnetic tape [1], with the tapes shipped to a central processing site when the observations were over. Tapes have now been replaced in most VLBI systems by inexpensive exchangeable computer disks. The Mk5A system, developed by MIT Haystack observatory [2] and used extensively by VLBI observatories in Europe and the US, uses a standard PC equipped with an 8-disk exchangeable disk pack and a commercial data streaming card. The maximum data rates have increased over the years, so that it is now possible to record almost continuously at 1024 Mbps. (Note that data rates in VLBI systems are in powers of 2 due to technical constraints.)

The European VLBI Network (EVN) is the organization that administers VLBI operations primarily within Europe, but it also includes telescopes in the Caribbean, Asia and South Africa. The EVN members operate 18 individual antennas, including some of the largest and most sensitive radio telescopes in the world [3], see figure (1).

(Fig 1 near here)

The recorded data are eventually shipped to a central processor, where the tapes or disks are played back together, and the signals cross-correlated. (In the case of European VLBI, this correlator is at the Joint Institute for VLBI in Europe (JIVE) in Dwingeloo, The Netherlands). The angular structure of radio sources can be obtained by Fourier inversion of the correlated data (the 'fringes') taken over a number of baselines. The number of Fourier components is effectively increased (and hence improving image fidelity) by means of Earth rotation, where the projected baseline length varies with time, tracing out part of an ellipse over ~12 hours [4]. The angular resolution is inversely proportional to the maximum length of the baseline, so trans-world arrays produce images showing the finest detail.

The ESLEA project [5] has enabled the use of the Internet for VLBI data transfer in the UK, which together with developments within the EVN has produced the first regularly operating real-time astronomical VLBI system in the world. The eVLBI subproject has been led by the team at The University of Manchester's Jodrell Bank Observatory (JBO).

## 2. eVLBI

The advent of high capacity international data networks has enabled a revolution in the way that VLBI observations take place. Tests in the 1990s over academic production networks showed that real-time transmission rates of a few Mbps were possible, where short sections of data from telescopes were transferred by FTP in order to check operations. However the big breakthrough came in 2002, when we were able to demonstrate data rates (producing correlation fringes) of 500 Mbps [6]. The development of a real-time VLBI system: eVLBI, then became a possibility, making wide use of the Internet.

Initial tests were at relatively low data rates, up to 32 Mbps, using the production academic network packet switched connections from telescopes to JIVE, but nevertheless interesting results were obtained [7]. eVLBI has a number of advantages, in particular the ability to check that everything is working immediately rather than the wait of several weeks or even months for tapes to be shipped and correlated. This can result in a significant increase in reliability for VLBI operations. In addition, the technique lends itself to rapid reaction observations – where subsequent observations on rapidly varying objects can be decided upon within hours rather than weeks after initial measurements are made.

The signal to noise ratio for a wideband continuum source in radio astronomy is proportional to the square root of the product of bandwidth and integration time, so bandwidth is as important as time. There is an obvious need for high data rates. Note that typically a data rate of 512 Mbps corresponds to a bandwidth of 128MHz, using 2 bit digitization (possible due to the characteristics of random signals) and a factor of 2 for the Nyquist rate. Very high data rates are in principle possible using fibre-optic technology, well in excess of that accessible in recording techniques, so eVLBI could eventually provide a much better sensitivity in high resolution astronomy. However the data needs to be continuously streamed, and usage of standard production networks is now such that congestion is likely to occur at rates of a few hundred Mbps.

Protocols such as TCP will drop data rates dramatically even with only one packet lost [8], so though the data loss may be negligible, the drop in data rate will seriously degrade the sensitivity of the array. The effects can be made worse by the long recovery time of long links running at high rates. The answer is to use dedicated lightpaths, where bandwidth is more easily guaranteed.

**3. The Network and ESLEA**

The GÉANT network [9] provides trans-national connections for academic users in the European Community. Individual countries have their own organisation for providing network services (National Research and Education Networks - NRENs). Standard production internet connections for Universities in the United Kingdom utilise the academic production network, Super JANET (SJ4 when this project started, now SJ5) [10], which provides internet and network communication for all the academic institutes throughout the UK. Universities and colleges throughout the Liverpool-Manchester corridor are connected to Net North West, which then joins Super JANET through a 2.5 Gbps link (now replaced by a 10 Gbps system in SJ5) in Manchester. For eVLBI, Super JANET is connected to the GÉANT network in London and then to SURFnet [11] (The Netherlands' academic network) for onward connection to Dwingeloo in the north of the Netherlands. Other NRENs are involved to connect other telescopes in Europe (currently experiments with up to 6 telescopes are possible). Figure (2) shows the telescopes in EVN currently connected to JIVE for eVLBI operations.

(Figure 2 near here)

The production network is a shared IP network and so transmission of VLBI data can be affected (quite drastically in some circumstances) by network congestion and packet loss within network routers. This reduces the available data rate, and long recovery times occur for TCP transfer over long links.

Dedicated optical networks provide a direct optical fibre link between two hosts. This means that network congestion and router loss should not be a problem. UKLight [12], a national R&D optical network of dedicated lightpaths operated by UKERNA [10], comprises multiple nodes throughout the UK and can peer with the international optical test bed. ESLEA [5] was the first widely scoped project to utilise UKLight.

ESLEA is a UK funded project designed to exploit dedicated lightpath technologies for use with many physics and engineering applications. UKLight can be used to transfer VLBI data to the correlator at JIVE using its peering ability with SURFnet [11], or indeed NETHERLight, an optical network in The Netherlands. The lightpaths needed for the applications in ESLEA were set up by the project team in collaboration with staff from UKERNA in liason with staff from the continental networks and JIVE.

(Figure 3 near here)

Figure 3 shows the connection between Jodrell Bank Observatory, situated 30 km South of Manchester, through UKLight, GÉANT and SURFnet to JIVE, used for tests and eVLBI experiments. The link to the site at Jodrell Bank was installed in

December 2004 and consists of a dedicated fibre pair carrying a 2.5 Gbps CWDM signal. By the end of the project the links enabled 1x 1 Gbps and 1x 0.6 Gbps connections to JIVE from Jodrell Bank, plus a 1 Gbps connection to Chicago. In theory, the network performance of dedicated lightpath networks such as UKLight and NETHERLight should be superior to production networks. The utilisation of these networks is essential in the development of real-time VLBI experiments.

**4. Development and testing for eVLBI**

The main objectives of the eVLBI part of the ESLEA project were to 1) demonstrate advantages of eVLBI data transport on UKLight over that available via production networks, 2) formulate methodologies for optimum use of optical networks for VLBI and 3) bring eVLBI to the user.

A two-fold approach to the required development was undertaken in order to achieve these objectives:
1. A series of tests using software tools on high performance computers and existing VLBI PCs was made to test the network and end hosts
2. A regular series of eVLBI tests initiated by the EVN using radio astronomy signals and Mk5A recording and playback systems at observatories and at JIVE.

Initial tests were made from Manchester. However optical fibres installed for e-MERLIN use became available in December 2004, enabling the connection of Jodrell Bank to be used in the testing.

**5. Network Performance Tests**

These involved the use of software designed to measure network performance. Since the full capacity of the links was 1 Gbps, high performance server-class PCs equipped with good quality network interface cards were required in order that the link performance was not limited by the end hosts. The two main protocols used in networks are Transmission Control Protocol over Internet Protocol (TCP/IP) and User Datagram Protocol (UDP). The programmes Iperf and BWCTL were used extensively to test TCP performance and UDPmon was used for UDP tests [13].

**5.1 UDPmon tests**

The first test was to prove that the test machines were powerful enough for the network tests to be undertaken. Two machines were connected back-to-back in the laboratory via network interface cards using 1 Gigabit Ethernet (1 GE). Figure 4 shows the throughput measured by UDPmon. The plot can be split up into three regions; A, B and C. The simplest of these regions is in region A, and is purely the result of an inverse power law. This effect arises from the fact that when the packet spacing is doubled, the throughput is halved. The second region, B, arises due to the fact that a packet exists for a finite length of time; the larger the packet the longer the length of time. This produces a lower limit on the packet spacing, below which it is simply not possible to transmit another packet, although it is possible to request spacing smaller than this limit. Thus, the x-axis of the plot shows the requested packet spacing, not the actual packet spacing. The third region, C, arises due to insufficient processing power in the sending or receiving machine. The plot shows typical UDPmon behaviour, and that the CPUs and network cars used are more than adequate when used with the larger (1472 bytes) packet size. Note that the actual packet sizes are larger due to the presence of header information (1500 bytes for the larger packets).

(Figure 4 near here)

The next series of tests were designed to compare production network connections with UKLight connections. Initially the connections from Jodrell Bank were routed onto the standard production network service at Manchester. Later the connection was changed to a lightpath, by setting up subnets with the known IP numbers of switches and routers in the path. A VLAN was formed between the end hosts. UDPmon (memory to memory) tests from Manchester to JIVE on UKLight and production links (see figure 5) clearly showed the advantages offered by dedicated lightpath connections. Small packets require more computing and network resources, and this results in a few lost packets (middle plot). The lower plot however shows that significant reordering has taken place. It is significant that tests on UKLight showed no reordering. Indeed traffic on Net North West is such that high bandwidth eVLBI on the production network would be impossible without causing serious disruption to other users.

(Figure 5 near here)

**5.2 TCP tests**

Tests were also made using the TCP protocol on the production and dedicated lightpath networks. These measurements make use of BWCTL which uses the Iperf software package to test the TCP performance of the network in memory to memory tests. Table 1 summarises the results on the available Mk5A equipment as well as high performance machines (gig 1 etc.). The results show that in most cases line rate (~980 Mbps) or close to it can be achieved. The exceptions are:

1. It is apparent that the transmission speeds between the Mk5A machines are noticeably slower than the transmission speeds between the high specification network PCs such as JBGig1, JIVEGig1, Gig7 and Gig8. This is not surprising since they are generally Pentium III machines with 1 GHz CPUs.
2. Rates of up to 800 Mbps could be achieved from JBO to JIVE. The JBO Mk5A equipment could not reach rates of 512 Mbps in eVLBI runs, whereas Onsala's Mk5A could.

It became clear from this that something was unexpectedly wrong with the JBO Mk5A equipment. It was also noteworthy that the maximum rate obtained using the JBO equipment was lower (800 Mbps) than that achievable from Onsala (980 Mbps) as shown in table 1. It was clear that the JBO Mk5A CPU was close to the limit - as shown in figure 6, where line rate was not achieved for any packets smaller than the maximum in UDPmon measurements. For a single TCP stream running at 777 Mbps from JBO to JIVE, CPU usage was found to be 96.3% in kernel mode with only 0.06 % in idle, whereas a shorter link JBO to Manchester) could run at 950 Mbps with 94.7% in kernel mode and 1.5 % idle. Adding a CPU intensive task reduced the throughput rate on JBO-Manchester, such that the rate was inversely proportional to the 'nice' parameter (figure 7). It was clear that the CPU power was insufficient for the desired task.

(Figure 6 near here)
(Figure 7 near here)

A higher performance motherboard (Intel Xeon 2.8 GHz processor) was installed on the JBO machine and we then found that UDPmon tests produced line rate (980Mbps) transfer routinely and that the desired 512 Mbps rate for eVLBI data transfer was obtained. As a result other machines in the network have now been upgraded, in particular the receiving computers in JIVE. We can only conclude that the Onsala machine had a slightly higher performance than that at Jodrell – just enough to raise the TCP eVLBI rate to 512 Mbps.

**Table 1. The transmission properties of the participating EVN telescopes and institutes.**

| Test Link | Network | Test | Machines | Data Rate Mbps |
|---|---|---|---|---|
| **JBO – JIVE** | Production | BWCTL | JBGig1 - JIVEGig1 | 980 |
| | | BWCTL | JBGig1 – Mark 5(575) | 900 |
| | | BWCTL | Mark 5(554) - JIVEGig1 | 600-800 |
| | | BWCTL | Mark 5(554) – Mark 5(575) | 600-800 |
| | | UDPmon | JBGig1 – JIVEGig1 | 980 |
| | | UDPmon | JBGig1 – Mark 5(575) | 900 |
| | | eVLBI | Mark 5(606) – Mark 5(575) | 504 |
| | Switched-Lightpath | BWCTL | JBGig1 - JIVEGig1 | 980 |
| | | BWCTL | JBGig1 – Mark 5(550) | 900 |
| | | BWCTL | Mark 5(606) - JIVEGig1 | 600-800 |
| | | BWCTL | Mark 5(606) – Mark 5(550) | 600-800 |
| | | UDPmon | JBGig1 – JIVEGig1 | 900 |
| | | eVLBI | Mark 5(606) – Mark 5(550) | 490 |
| **Man – JIVE** | Production | BWCTL | Gig7 – JIVEGig1 | 980 |
| | | BWCTL | Gig7 – Mark 5(575) | 900 |
| | | UDPmon | Gig7 – JIVEGig1 | 980 |
| | | UDPmon | Gig7 – Mark 5(575) | 900 |
| | Switched - Lightpath | BWCTL | Gig8 – JIVEGig1 | 980 |
| | | BWCTL | Gig8 – Mark 5(550) | 900 |
| | | UDPmon | Gig8 – JIVEGig1 | 980 |
| | | UDPmon | Gig8 – Mark 5(550) | 900 |
| **Onsala – JIVE** | Production | BWCTL | ON_Mark 5 – Mark 5(524) | 980 |
| | | eVLBI | ON_Mark 5 – Mark 5(524) | 512 |
| **Westerbork – JIVE** | Dark Fibre | BWCTL | Mark 5(586)– Mark 5(608) | 980 |
| | | eVLBI | Mark 5(586)– Mark 5(608) | 512 |
| **Torun – JIVE** | Production | eVLBI | Mark 5(611)– Mark 5(620) | 256 |
| **Medicina – JIVE** | Production | eVLBI | ME_Mark 5– Mark 5(558) | 128 |

## 6. Multiple Data Flows

TCP achieves bit-wise correct data transfer, though the time of arrival is not guaranteed. On the other hand eVLBI and also some video streaming applications have constant bit rate, and the timeliness of the data arrival is as important as data fidelity [8]. In fact VLBI systems are designed to cope with high (up to ~10%) packet loss [14] without loss of synchronisation since they were originally designed to be used with magnetic tape. Software (VLBI_UDP) to transfer data at constant bit rate, using UDP has been developed for use in tests and eventually in eVLBI systems [15]. eVLBI will operate with data streams from many telescopes travelling to JIVE over the various networks involved. A test of multiple data streams was therefore initiated, using VLBI_UDP to transfer continuous flow data in memory to memory tests.

(Figure 8 near here)
(Figure 9 near here)

Figure 8 shows throughput for flows from Jodrell Bank, Medicina (Italy) and Torun (Poland) observatories into JIVE, obtained during a demonstration at the GÉANT2 launch in June 2005. The unusual nature of the periodic drops in throughput seen on all three flows made us suspicious of this experiment, which we planned to repeat. However, owing to hardware problems, the repeat tests did not happen until after further development of VLBI_UDP. Figure 9 shows a test made in December 2006. Throughputs of 800 Mbps were obtained on the Manchester-JIVE lightpath, and 600 Mbps and 400 Mbps obtained on the packet switched connections from Manchester and Bologna respectively. High (up to 0.6%) packet loss was seen on the Manchester-JIVE production link. This was due to temporary bonding between 1 Gbps links, all being used at 50% during an upgrade at Manchester to Super JANET 5.

**7. Trans-Atlantic data flows**

iGRID 2005 and Super Computing 2005 (SC2005) gave us an opportunity to test the performance of dedicated lightpaths over long links. Telescopes in the USA regularly join in with the EVN in global VLBI experiments, where the extra resolution given by trans-Atlantic baselines is needed. eVLBI tests were therefore organised where data from European telescopes (Onsala and Westerbork) were sent across GÉANT to London, UKLight lightpaths to Chicago and then via the HOPI and DRAGON networks to Haystack observatory in Massachusetts. Data from Jodrell Bank, Onsala and Westerbork were successfully transferred at line rates using three VC-3-13c SDH circuits on trans-Atlantic cable, and an eVLBI experiment was successfully conducted between Onsala and a telescope at Haystack at a rate of 512 Mbps. Dynamic provisioning under GMPLS was also tested during this experiment [16].

The link to Haystack was dismantled after SC 2005; however a 1 Gbps lightpath from Manchester to Chicago was established over UKLight early in 2007. Tests between server class machines at Jodrell Bank and at the Chicago Starlight Point of Presence using UDPmon and Iperf however showed intermittent packet loss. Further tests after maintenance work on the links by UKERNA showed outstanding performance, with only 4 packets lost in $2 \times 10^{10}$ transmitted continuously over 2.5 days at 940 Mbps [17]. This test demonstrated the need for connectivity testing before applications are run. Trans-Atlantic eVLBI runs are planned for the near future.

**8. eVLBI tests**

The EVN organised a series of tests throughout the project involving an increasing number of telescopes in the network as they became fitted out with high bandwidth internet connections. The tests were roughly every 6 weeks, fitted around regular VLBI observing sessions. The prime aim was to develop the bandwidth capability and reliability of the network, using the Mk5A VLBI hardware. The following list demonstrates how the technology has developed over recent years. The items where ESLEA and the use of lightpaths provided under the ESLEA project have made a major contribution are shown in italics.

- January 2004: A disk buffered eVLBI session with 3 telescopes at 128Mb/s achieved the first eVLBI image
- April 2004: A 3 telescope real-time eVLBI session, obtained fringes at 64Mb/s: and produced the first real-time EVN image at 32Mb/s.

- September 2004: A 4 telescope real-time eVLBI run obtained fringes to Torun and Arecibo and formed the first EVN, eVLBI science session to ;produce astronomical results
- *December 2004: Connection of JBO to Manchester by 2 x 1 Gbps links and, eVLBI tests between Poland, Sweden, the UK and the Netherlands at 256 Mb/s*
- January 2005: Australia to JIVE: data from the descent of the Huygens probe to the surface of Saturn's moon Titan transferred at ~450Mb/s
- February 2005: Haystack (US) – Onsala (Sweden) runs at 256 Mb/s
- *2005: Regular tests with VLBI data every ~6 weeks,128 Mpbs worked reliably,, 256 Mpbs often, Onsala can run at 512 Mbps but not JBO*
- *June 2005: GÉANT2 launch, 3 way flows up to 800 Mbps*
- *Sept/Nov 2005: iGRID 2005, SC 2005 – Trans-Atlantic data transfer Onsala, JBO, Westerbork to Haystack correlator – real time eVLBI at 512 Mbps*
- *Feb 2006: Upgrade to JBO Mk5A CPU – 512 Mbps now possible*
- *April 2006: First successful open call EVN eVLBI session*
- *Sept 2006: Packet dropping tests to JIVE correlator*
- *Dec 2006: repeated 3-way flows on GÉANT2*
- *2007: Regular science sessions in operation*

It can be seen that the tests have resulted in the setting up of routine eVLBI astronomy sessions and have effectively brought eVLBI to the user. Two astronomical publications have already resulted from this work [18,19].

## 9. Conclusion and Future Work

In summary, the ESLEA e-VLBI project has been very successful, achieving all its original aims, and even leading to further developments. We are now able to perform routine real time eVLBI measurements at data rates of 256 Mbps, and 512 Mbps observations on all 6 telescopes currently connected to JIVE will shortly be brought into operation. This level of performance is impossible to achieve using the production network, where in some cases (e.g. from Manchester) there is just not sufficient capacity available on the data links.

The availability of lightpaths has enabled 24-hour observing sessions open to all astronomers to be run every 6 weeks. An illustration of such an observing run occurred when the first rapid response experiment was undertaken (investigators A. Rushton and R. Spencer). Here a 6 telescope real time observation was run on 29th Jan 2007 at 256 Mbps. The results were analysed overnight as data came from the correlator in a fraction of the normal time taken in VLBI in order to select celestial radio sources for follow up observations on 1st Feb. This kind of operation would be impossible for conventional VLBI. The experiment worked well technically - 16 sources (weak microquasars) were successfully observed, but all were <0.5 mJy – too weak to observe in the follow up run – indicating a perverse universe. The feasibility of the technique was clearly demonstrated. The main delay now is in refereeing of the proposal, rather than technical issues to do with the observations and processing. A new class of triggered observations where a proposal can be vetted well before the allocated dates has been initiated, where the actual trigger can come within hours of the observing run.

It is clear that there is a future in eVLBI, and so the members of the EVN have joined in a project to provide links to all the EVN telescopes. This EU funded project, EXPReS, will provide the infrastructure for these links, connect multiple telescopes of the UK's eMERLIN array into the network, support distributed processing using GRID methodologies and engage in the development of 4 Gbps systems. This project will create a distributed, large-scale astronomical instrument of continental and inter-continental dimensions.

**References**

[1] Moran, J., Very Long Baseline Interferometer Systems, in Methods of Experimental Physics: Astrophysics, Part C, ed. M. L. Meeks, New York, Academic Press, 12, 174, 1976

[2] MIT Haystack Observatory: Mk5, viewed 23 May 2007, http://www.haystack.edu/tech/vlbi/mark5/index.html

[3] European VLBI Network (EVN), viewed 23 May 2007, http://www.evlbi.org/

[4] Thompson, A., Moran, J., and Swenson, G., Interferometry and Synthesis in Radio Astronomy, 2nd ed., Wiley-VCH, 2001

[5] ESLEA Web Page, viewed 23 May 2007, http://www.eslea.uklight.ac.uk/

[6] Hughes-Jones, R., Parsley, S., Spencer, R., "High data rate transmission in high resolution radio astronomy – vlbiGRID", Future Generation Comp. Syst. 19(6), 883-896, 2003

[7] JIVE, First Science with e-VLBI, viewed 23 May 2007 http://www.jive.nl/news/first_science/first_science.html

[8] Kershaw, S. et al., "TCP delay: Constant bit rate data transfer over TCP", in "Lighting the Blue Touch Paper for UK e-Science – Closing Conference of ESLEA Project", PoS(ESLEA)002, 2007. See http://pos.sissa.it

[9] GEANT web page, viewed 23 May 2007, http://www.geant.net/

[10] JANET and UKERNA, viewed 23 May 2007, http://www.ukerna.ac.uk/

[11] SURFnet web site, viewed 23 May 2007, http://www.SURFnet.nl/info/en/home.jsp

[12] UKLight web site: http://www.uklight.ac.uk/

[13] Richard Hughes-Jones web site, viewed 23 May 2007 http://hep.man.ac.uk/u/rich/

[14] Casey, S. et al. "Investigating the effects of missing data upon VLBI correlation using the VLBI_UDP application", in "Lighting the Blue Touch Paper for UK e-Science – Closing Conference of ESLEA Project", PoS(ESLEA)025, 2007.  See http:/pos.sissa.it

[15] Casey, S., et al. « VLBI_UDP", in "Lighting the Blue Touch Paper for UK e-Science – Closing Conference of ESLEA Project", PoS(ESLEA)038, 2007.  See http:/pos.sissa.it

[16] Sobieski, J., Lehman, T., Jabbari, B., Ruszczyk, C.,  Summerhill, R., Whitney, A.,  "LightPath services for radio astronomy applications", Future Generation Comp. Syst. 22(8): 984-992, 2006

[17] Rushton, A., Burgess, P., Hughes-Jones, R., Kershaw, S., Spencer, R., Strong, M., "Trans-Atlantic UDP and TCP network tests" in "Lighting the Blue Touch Paper for UK e-Science - – Closing Conference of ESLEA Project", PoS(ESLEA)007, 2007.  See http:/pos.sissa.it

[18] Rushton, R., Spencer, R. E.,  Strong, M.; Campbell, R. M., Casey, S., Fender, R. P., Garrett, M. A., Miller-Jones, J. C. A.; Pooley, G. G., Reynolds, C., Szomoru, A., Tudose, V. Paragi, Z., "First e-VLBI observations of GRS1915+105", MNRAS 374, L47, 2007

[19] Tudose, V. Fender, R. P. Garrett, M. A. Miller-Jones, J. C. A., Paragi, Z., Spencer, R. E., Pooley, G. G., van der Klis, M., Szomoru, A., "First e-VLBI observations of Cygnus X-3", MNRAS, 375, L11, 2007

**Acknowledgements**

The authors would like to thank Arpad Szomoru and the staff at JIVE and at EVN observatories for their patience, forbearance and invaluable efforts in making eVLBI a reality. EXPReS is an Integrated Infrastructure Initiative (I3), funded under the European Commission's Sixth Framework Programme contract number 026642 EXPReS.

**Figure captions:**

Figure 1. Map of the globe showing the positions of member telescopes of the European VLBI network (EVN).

Figure 2. Telescopes and  networks used in eVLBI operations in May 2007. It is planned to include all the EVN telescopes once fibre connection have been made, as part of the EXPReS project.

Figure 3. Diagram showing the switched-lightpath network connections between Jodrell Bank and JIVE

Figure 4. Back to back throughput of two server class machines showing the typical results of UDPmon tests. Note that the horizontal axes in this figure and in Figure 5 are in units of microseconds. The curves in different colours show how the data rate varies with demanded inter-packet spacing for a variety of packet sizes available to the user. Most networks allow total packet sizes of 1500 bytes, so the top (light blue) curve is the most relevant to expected performance.

Figure 5. UDPmon tests on the production academic network between Manchester and JIVE in the Netherlands. The plot shows throughput (upper), packet loss (middle) and packet reordering (lower) versus demanded inter-packet spacing for a variety of packet sizes. The bottom curve shows that significant packet reordering took place, which was not seen between the same machines connected via a UKLight lightpath.

Figure 6. UDPmon test between a Mk5A VLBI machines at Jodrell Bank and a similar Machine at JIVE. Throughput drops significantly for packets smaller than the maximum, with a concomitant increase in packet loss. These results should be compared with those on figure 5.

Figure 7. The top graph shows the data throughput from Jodrell Bank to Manchester over the dedicated optical link as a function of the priority assigned to an additional CPU intensive task (the nice parameter in Linux). A higher nice parameter implies a lower priority, allowing more resource to be devoted to data transfer. The middle graph shows the CPU usage when transferring data between Jodrell Bank and Manchester over the dedicated optical link as a function of the nice priority of the additional CPU intensive task. Bottom: graph showing throughput of data from Jodrell Bank to Manchester over the dedicated optical network as a function of the available CPU in Kernel mode.

Figure 8. Plots of throughput for VLBI_UDP multiple simultaneous data flows across GÉANT2 over a 5-hour period in June 2005. Upper: flow between JBO to JIVE from a 2.0 GHz dual Xeon machine to a 2.4 GHz dual Xeon. Rates of 670-840 Mbps were achieved. Middle: Medicina (Bologna) to JIVE from an 800 MHz PIII to a Mk5 (623) 1.2 GHz PIII at JIVE. The rate, 330 Mbps, was limited by the sending PC. Lower: Toruń to JIVE using a 2.4 GHz dual Xeon to a Mk5 (575) 1.2 GHz PIII. The maximum rate of 325 Mbps here was limited by security policing. The ~17 minute period in throughput (accompanied by concomitant changes in packet loss) was likely to be caused by congestion in the links. It did not appear in subsequent tests. 1500 byte packets were used throughout.

Figure 9. Plots of 3-way continuous data flows using VLBI_UDP, between Manchester and JIVE on both a dedicated lightpath on UKLight and the production network, together with a flow from Bologna (nearest point of presence to the telescope at Medicina) and JIVE. The dark blue curves show throughput, the lighter, pink curves the percentage of lost packets.

Fig 1

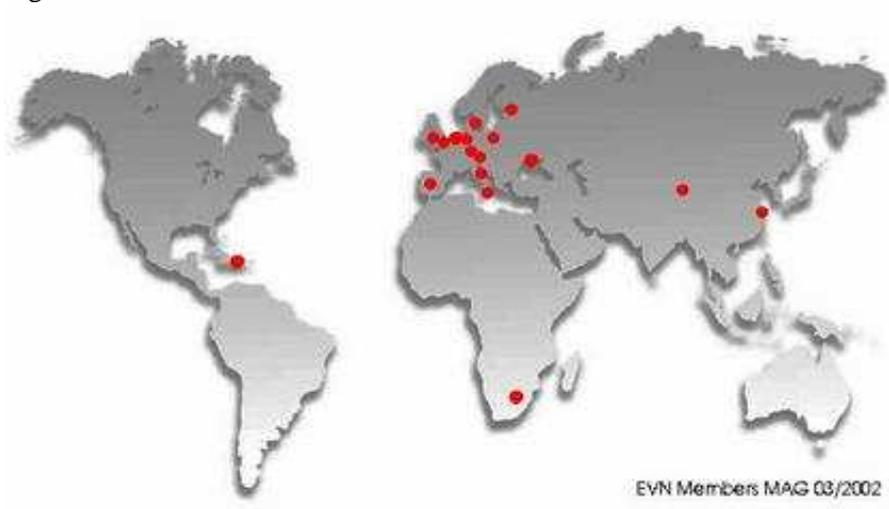

Fig 2

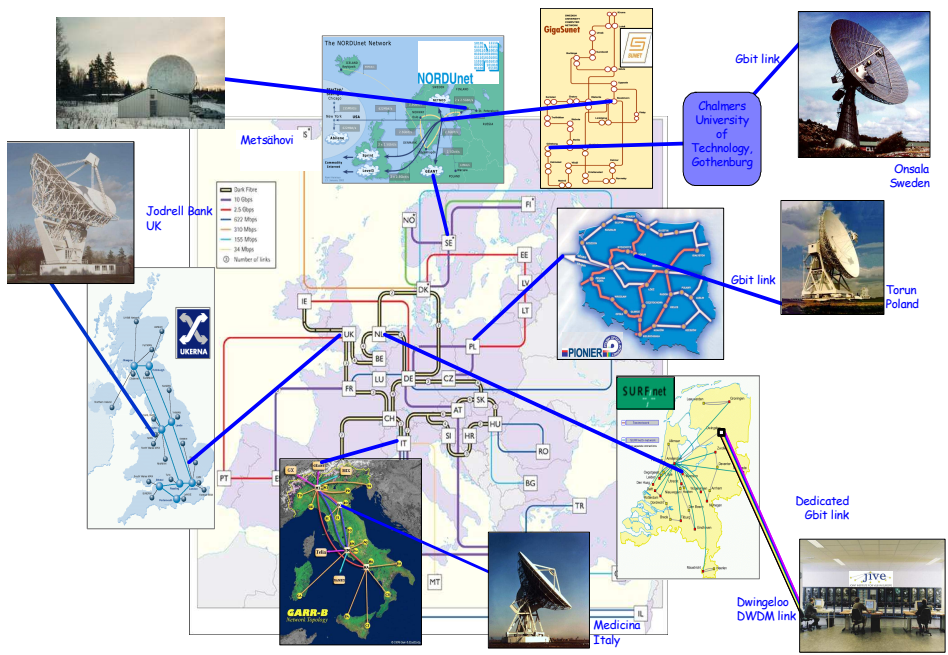

Fig 3

Fig 4

Fig 5

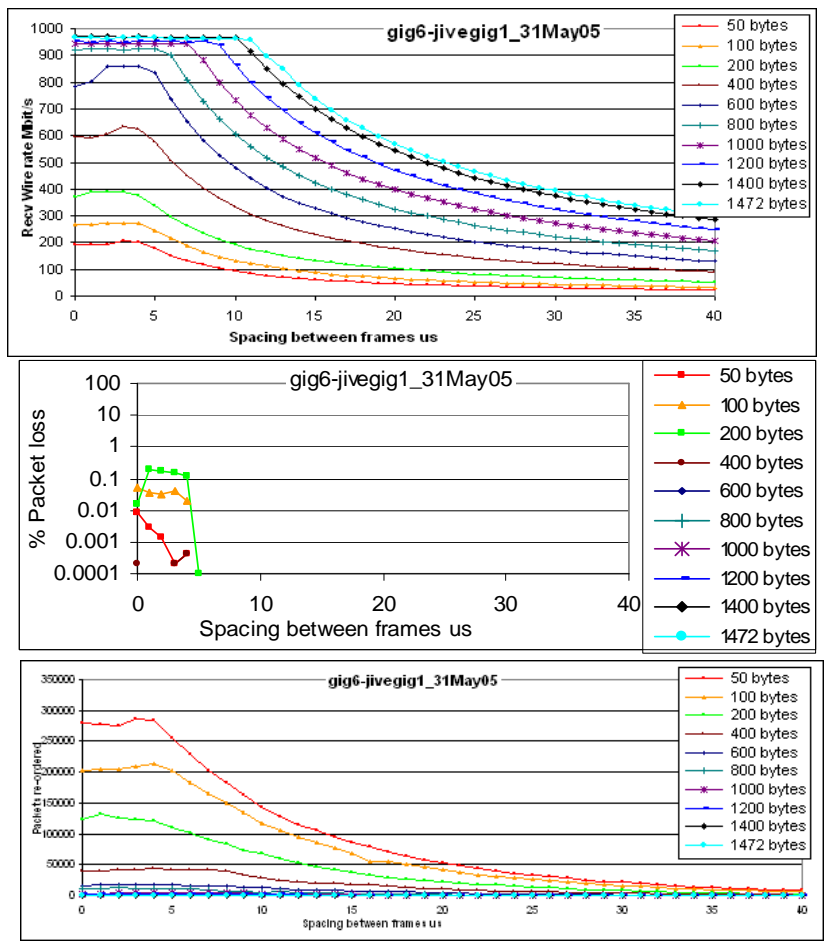

Fig 6

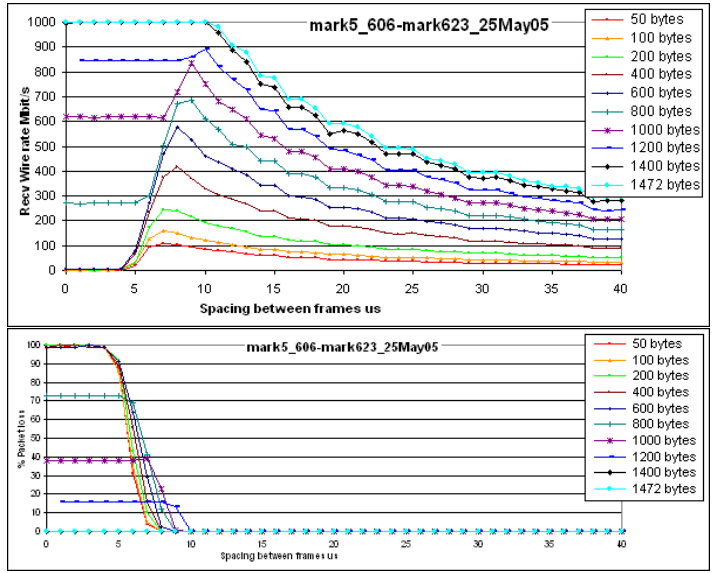

Fig 7

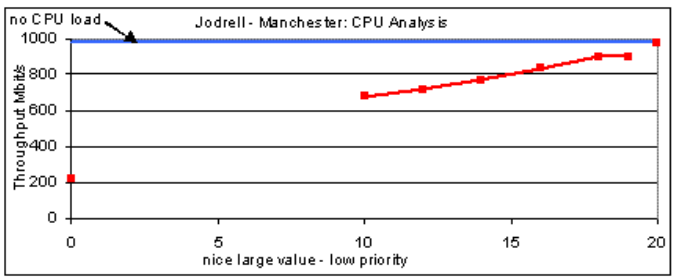

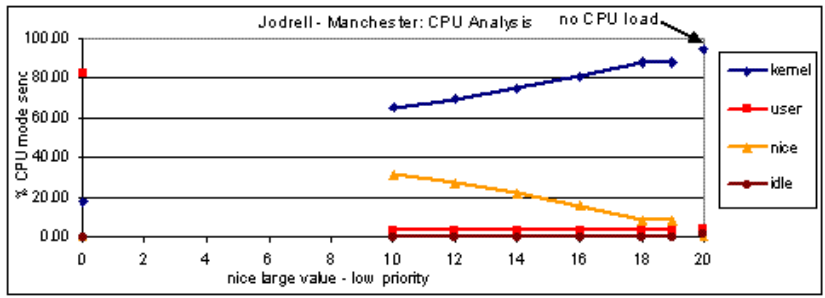

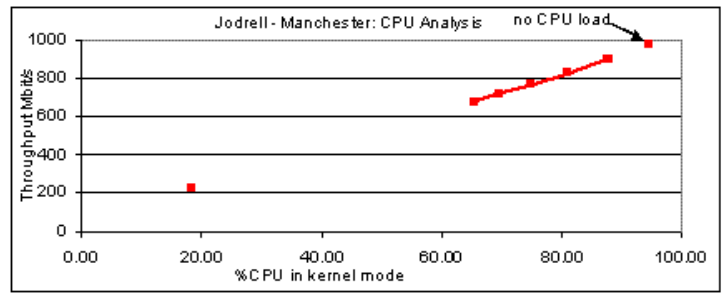

Fig 8

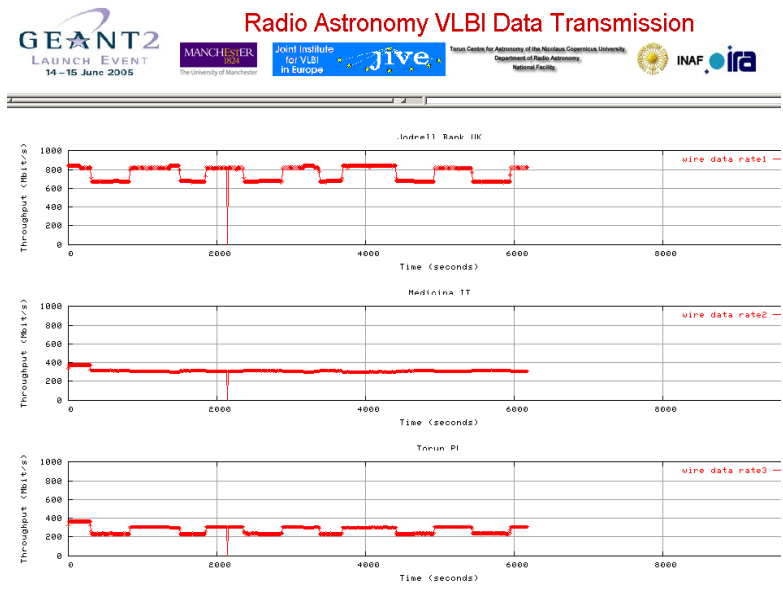

Figure 9

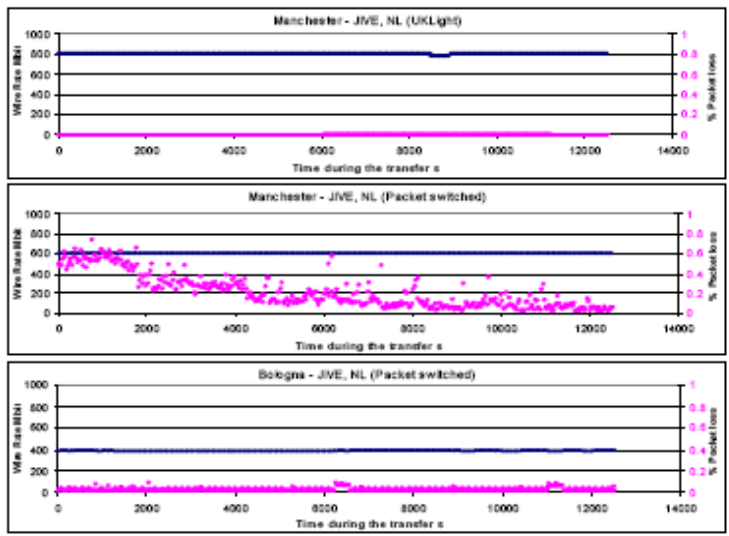

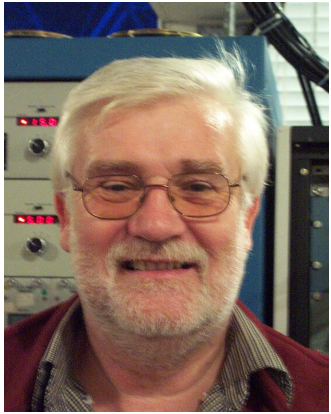

Dr Ralph Spencer is Reader in Radio Astronomy at the University of Manchester. He has worked on the development of interferometers for radio astronomy since the 1970's, leading work on the development of phase stable radio linked interferometers with 100 km baselines. This work led to the development of the MERLIN array of 7 telescopes now operated as a National Facility by the University of Manchester for STFC. His interest in VLBI started in the late 1970's and he was responsible for running VLBI operations at Jodrell Bank Observatory until the mid 1990's, when the National Facility took over. European development of the 1 Gbps MkIV VLBI system was led by him until 2001, and this is now in routine operation world wide at data rates up to 256 Mbps. Recently he has lead the group at Jodrell Bank developing eVLBI systems. His astronomical research interests include studies of microquasars. These are radio emitting X-ray binaries in our own galaxy which emit jets of matter at relativistic velocities. They act as laboratories for the study of the formation of jets in general, a topic which is great relevance to the study of quasars and radio galaxies in general. Ralph is a Fellow of the Royal Astronomical Society.

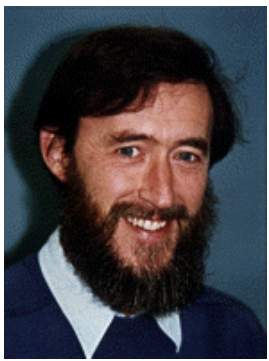

Dr Richard Hughes-Jones leads the e-science Grid Network Research and Development in the Particle Physics group at Manchester University. He has a PhD in

Particle Physics and has worked on Data Acquisition and Network projects. He is a member of the Trigger/DAQ group of the ATLAS experiment in the LHC programme, focusing on Gigabit Ethernet, protocol performance, application gateways, and remote computing farms. He has been responsible for the High performance, High Throughput network investigations in the European Union DataGrid and DataTAG projects, the UK e-Science MB-NG, and GridPP projects. Currently he is a co-PI of the UK e-Science ESLEA project focusing on delivering high performance networking using switched lightpaths to science users including VLBI, HEP as well as other e-Science users. He is also a member of the EU EXPReS project working on protocol and high-bandwidth network performance for future VLBI. Richard is an area director for the Open Grid Forum and a co-chair of the Network Measurements Working. He is also a co-chair of PFLDnet 2005 & 2006, and 2008 and a program committee member of the IEEE Real Time Conferences. He has recently been a contributor to the book "Grid Networks: Enabling Grids with Advanced Communication Technology".

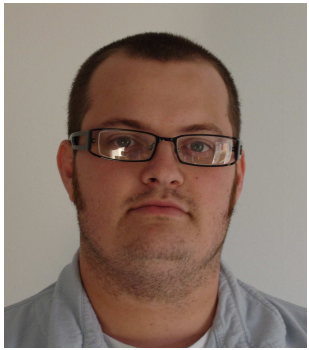

Dr Matthew Strong is currently a Project Physicist working on sensors for industrial applications. He obtained a MSci (Hons) degree in Physics and Astronomy at the University of Durham with emphasis on optical fibres for astronomy. He continued his education obtaining a PhD in Astronomy and Astronomical Technology at The University of Manchester, during which time he worked on optical fibre technologies for radio interferometers. He also led astronomical research projects making use of various radio astronomical interferometers. Matthew's interests in astronomical technologies and astronomical research areas such as Active Galaxies led him to accept an ESLEA PDRA position at Jodrell Bank to work on high speed data transmissions and the development of e-VLBI. Matthew is a Fellow of the Royal Astronomical Society and a Member of the Institute of Physics.

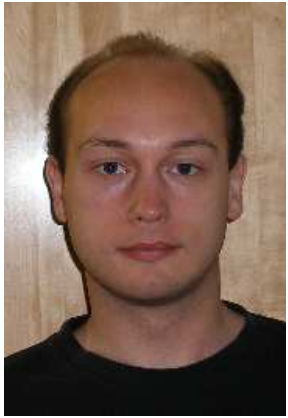

Simon Casey is a third year PhD student in Radio Astronomy at the University of Manchester, researching the development of e-VLBI, in particular VLBI-UDP. He completed his undergraduate studies at the University of Manchester, attaining an upper second class MPhys degree in Physics with Technological Physics.

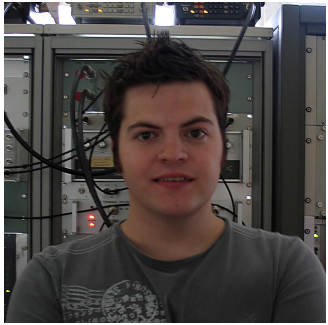

Anthony Rushton is studying for his Ph.D. at Jodrell Bank Observatory, The University of Manchester, funded by the STFC. In 2004 he became a Fellow of the Royal Astronomical Society and in 2005 he received a Masters degree in Physics from the University of Manchester. He is currently researching the radio emission of superluminal ejections from X-ray binaries using MERLIN and VLBI instruments. His research interests also include the application and development of eVLBI and high bandwidth network applications. eVLBI exploits the huge growth of the Internet to transfer large volumes of radio data across multiple continents. This allows the real time imaging of the radio sky at milliarcsecond resolutions, furthering our understanding of transient celestial phenomena.

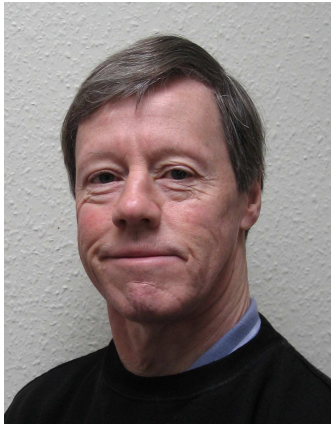

Paul Burgess is the operations manager for VLBI observations at Jodrell Bank. As well as being responsible for day to day running of VLBI experiments, he has taken a major part in the development of VLBI technology, including control software, wide band recording and lately eVLBI. Paul's first degree was in BSc (Hons) Physics and Electrical Engineering from the University of Manchester (1974) and he obtained an MSc in Instrument Design from the University of Manchester Institute of Science and Technology (UMIST) in 1983. He is a chartered member of the IoP and has had nine years experience in industrial electronics design followed by 23 years at Jodrell Bank.

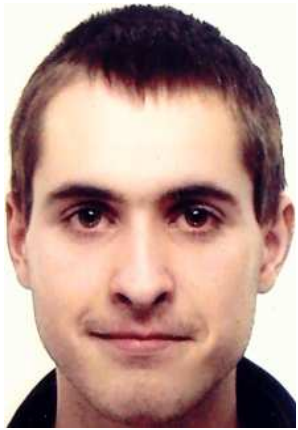

Stephen Kershaw is a Research Assistant at The University of Manchester, researching high-performance networks with Richard Hughes-Jones and working towards a PhD. He graduated from the University of Manchester in 2006 with a first-class Masters degree in Physics. His interests are in networks and protocols for eVLBI, multi-gigabit data transfer and in evaluating the performance of Linux systems.

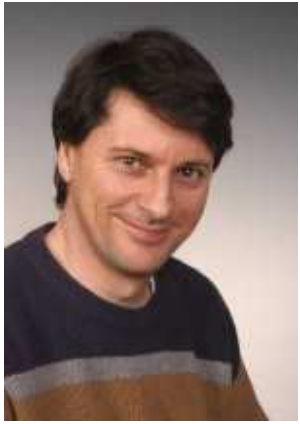

Colin Greenwood is Executive Officer at the International Square Kilometre Array Project Office, and EU-funded initiative to build the world's largest telescope, and is based at ASTRON, Netherlands. Previously, he was ESLEA Project Manager and e-Science Liaison Officer for the UK Government's DTI-funded Grid Computing Now! Knowledge Transfer Network, based at the National e-Science Centre (NeSC), Edinburgh University.  He has private sector experience of leading application integration projects for multi-national clients.   Other roles have included small business advisor, business development, research fellow and geologist.  Colin graduated with an MSc from Camborne School of Mines and with an MBA from Durham University Business School.  He is a Chartered Marketer and PRINCE2 practitioner.